\documentclass[aps,twocolumn,showpacs,floats,superscriptaddress]{revtex4}
\usepackage{graphicx}
\usepackage{amssymb,amsmath}
\usepackage{color}
\usepackage{dcolumn}
\usepackage{bm}

\def\he4{$^4$He}
\def\Am3{\AA$^{-3}$}
\def\beq{\begin{equation}}
\def\eeq{\end{equation}}
\begin{document}

%%%%%%%%%%%%%%%%%%%%%%%%%%%%%%%%%%%%%% AUTHORS %%%%%%%%%%%%%%%%%%%%%%%%%
\author{\c{S}.G. S\"{o}yler}
\affiliation{Department of Physics, University of Massachusetts,
Amherst, MA 01003, USA}

\author{A.B. Kuklov}
\affiliation{Department of Engineering Science and Physics,
CUNY, Staten Island, NY 10314, USA}

\author{L. Pollet}
\affiliation{Department of Physics, Harvard University, Cambridge, MA 02138, USA}

\author{N.V. Prokof'ev}
\affiliation{Department of Physics, University of Massachusetts,
Amherst, MA 01003, USA}
\affiliation{Russian Research Center ``Kurchatov Institute'',
123182 Moscow, Russia}

\author{B.V. Svistunov}
\affiliation{Department of Physics, University of Massachusetts,
Amherst, MA 01003, USA}
\affiliation{Russian Research Center ``Kurchatov Institute'',
123182 Moscow, Russia}

%%%%%%%%%%%%%%%%%%%%%%%%%%%%%%%%%%%%%%%%%%%%%%%%%%%%%%%%%%%%%%%%%%%%%%%%%%%%%%

%\title{Superclimb of  Dislocations and the Effect of Isochoric Compressibility in Solid \he4}
\title{Superclimb of  Dislocations and the Anomalous Isochoric Compressibility of Solid \he4}

%%%%%%%%%%%%%%%%%%%%%%%%%%%%%%%%%%%%%%%%%%%%%%%%%%%%%%%%%%%%%%%%%%%%%%%%%%%%%%
\date{\today}

\begin{abstract}
In the experiment on superfluid transport in solid \he4 [PRL {\bf 100}, 235301 (2008)], Ray and Hallock  observed an {\it anomalously large  isochoric compressibility}:  the supersolid samples demonstrated a significant and apparently spatially uniform response of density and pressure to chemical potential, applied locally through Vycor ``electrodes". We propose that the effect is due to {\it superclimb} : edge dislocations can climb because of mass transport along superfluid cores. We corroborate the scenario by {\it ab initio} simulations of an edge dislocation in solid \he4 at  $T=0.5K$. We argue that at low temperature the effect must be suppressed due to a crossover to the smooth dislocation.
\end{abstract}


% 67.80.bd Superfluidity in solid 4He, supersolid 4He
% 67.80.dj  Defects, impurities, and diffusion (in quantum solids)
% 67.80.-s  Quantum solids (generic)
% 05.30.Jp Boson systems (for static and dynamic properties of Bose?Einstein condensates, see 03.75.Hh and 03.75.Kk; see also 67.10.Ba Boson degeneracy in quantum fluids)

\maketitle

At present, the experimental search for supersolidity (proposed theoretically in Refs.~\cite{Andreev69})
in \he4 focuses mostly on torsional oscillator  experiments \cite{KC}, and on attempts to detect pressure driven  non-plastic flow \cite{meisel}.
So far, direct superflow through solid \he4 has been observed only in the  experiment by Ray and Hallock \cite{Ray,Ray2}. Although it was argued that
the presence of a few liquid channels is compatible with the observations~\cite{Balibar, Balibar_review}, the absence of flow at temperatures above $T \approx 0.6$K is a strong argument against this scenario and in favor of superfluidity along dislocation cores or grain boundaries. 
Theoretically, a superfluid dislocation network can manifest itself as a genuine superfluid or be in the Shevchenko state \cite{Shevchenko}, characterized by anomalously low viscosity due to phase slips. In practice, the Shevchenko state might mimic superfluidity even at relatively high temperatures,  $T > 0.1$K, well above the actual transition determined by the dislocation density.

The ``UMass sandwich" setup of Refs.~\cite{Ray,Ray2} is different from the pressure driven cells \cite{meisel}.
Superfluid \he4 is fed into the crystal through Vycor ``electrodes",
meaning that the chemical potential $\mu$ is the physical quantity relevant to the external perturbation applied to the crystal.
An insulating (i.e., non-supersolid) crystalline  groundstate has to be {\it  isochorically incompressible}:
$\chi\equiv (dn/d\mu)_V =0$; that is, the density $n$ of the crystal should demonstrate no response to infinitesimal, quasi-static changes of $\mu$ \cite{Mott}.
Indeed, as long as the creation of single vacancies  and interstitials is forbidden  by finite energy gaps, the only way the density of a crystal can react  dynamically to a small change in the chemical
potential, $\delta \mu$, is by creating/removing crystalline layers.
This requires nucleation times exponentially large in $|\delta \mu|^{-1}$. Thus, at temperatures much smaller than
the vacancy/interstitial gaps, the isochoric compressibility $\chi$ associated with thermally excited vacancies and interstitials is exponentially small.
Consistent with these arguments, all non-supersolid samples of Refs.~\cite{Ray,Ray2} have $\chi=0$: two pressure gauges monitoring the solid showed no response to a change in $\mu$ by the Vycor electrodes.

Supersolids have no vacancy (interstitial) gap \cite{PS} and are thus genuinely isochorically compressible: $\chi \neq 0$.
One might argue that  $\chi$ should scale linearly with the superfluid fraction $\rho_s$
since both are due to zero-point vacancies (interstitials). Given the extremely low value of $\rho_s\lesssim 10^{-5}$ following from  estimates based on the
observed supercritical flux (see Refs.~\cite{Ray,Ray2} for more details), one does not expect a noticeable $\chi$.
However, a density/pressure response to $\mu$ by several orders of magnitude larger than expected is precisely what was observed, to which we refer as the {\it effect of anomalous isochoric compressibility}. Remarkably, the response was apparently spatially homogeneous, since two pressure gauges attached to two ends of the solid typically showed equal
variations (but different absolute values; most samples were characterized
by a static pressure gradient) \cite{Ray,Ray2}.

In this Letter, we argue that the microscopic phenomenon behind the effect of anomalous isochoric compressibility in the experiments by Ray and Hallock is the {\it superclimb} of superfluid edge dislocations, that is, climb controlled by superfluid flow along the core. 
Our idea is that significant and spatially uniform mass accumulation in the bulk of supersolid \he4
is due to the synergy between: (i) the presence of a superfluid network
capable of delivering \he4 atoms from Vycor electrodes to distant bulk regions and (ii) the presence of edge dislocations, whose superclimb is responsible for the density/pressure change.

We corroborate our scenario by {\it ab initio} simulations which show that  edge dislocation with Burgers vector along the {\it hcp} C-axis has superfluid core  (we previously reported the superfluidity in the core of a screw dislocation \cite{Boninsegni07}), and that it can climb in response to variations of $\mu$. We argue that at low temperature the climb must be suppressed due to a crossover from a rough to a smooth dislocation~\cite{no_rough}. This prediction is a manifestation of the structural evolution of dislocations with temperature, and is important for experimental validation of the scenario. While superflow is a necessary condition for superclimb, the dislocation must also  have a finite density of jogs to allow for threshold-less climb. Otherwise, a finite gap $\Delta$ for creating dislocation jogs will protect the dislocation from shifting significantly in response to small variations in $\mu$.

The effect of anomalous isochoric compressibility is one of the novel properties
emerging in the ``quantum metallurgy" \cite{Dorsey} context.
These properties have long been discussed in the past; for example,
it was speculated that quantum dislocations should be characterized by ``thick" (roughened)
cores due to zero-point motion \cite{Meyerovich}. An important role of quantum roughening of dislocations in the torsional oscillator response has also been proposed in Refs.~\cite{deGennes,Biroli}.
Superclimb is a quantum analog of classical high-$T$ climb due to thermally activated flux of vacancies toward, away or along the cores (pipe diffusion) \cite{Hull} which adds (removes) atoms to (from) the extra plane forming the edge dislocation, so that the dislocation core shifts along the extra-plane direction. Obviously, at low $T$, the activated mass flow is exponentially suppressed and quickly becomes negligible. %In contrast, in solid \he4 the superflow along the core controls the climb.

%Our idea is that significant and spatially uniform mass accumulation in the bulk of supersolid \he4
%is due to the synergy between: (i) the presence of a superfluid network
%capable of delivering \he4 atoms from Vycor electrodes to distant bulk regions and (ii) the presence of edge dislocations, whose superclimb is responsible for the density/pressure change.
%First-principles simulations of solid \he4  have revealed superfluidity in the core of a screw dislocation \cite{Boninsegni07}.
%Here we report on the core superfluidity of the edge dislocation with Burgers vector along the {\it hcp} C-axis.
%We also find that this dislocation exhibits climb (assisted by its core superfluidity), which we refer to as
%{\it superclimb}.

Apart from climb, dislocations can also glide. In Ref.~\cite{no_rough} it was shown that gliding dislocations (gliding does not require mass influx) are smooth at $T=0$ because   Coulomb-type interactions between shape fluctuations \cite{Hirth,Kosevich} induce an energy gap $\Delta_{\rm glide}$ with respect to creating a pair of kinks in Peierls potential.  Hence, threshold-less glide of a dislocation can effectively occur only at $T$ comparable with $\Delta_{\rm glide}$. This gap is also related to shear modulus stiffening at low $T$ \cite{Beamish}. Similarly, dislocations have a gap $\Delta$ for creating a pair of jogs at $T=0$, which leads to a suppression of climb (and $\chi$) at low $T$. The values of $\Delta$ can be quite different from $\Delta_{\rm glide}$ because the jog--{\it anti}-jog deconfinement couples to fluctuations of the superfluid density leading to an additional mechanism for the gap formation.

{\it Model of climbing superfluid dislocation}. We introduce a coarse-grained description of an edge dislocation with superfluid core oriented along the X-axis in terms of  the core displacement $y(x, \tau)$ along the Y-axis (in the climbing direction), perpendicular to the Burgers vector which is along the Z-axis.  We proceed under the assumptions of small gradients and large displacements compared to the lattice spacing. Then,  a coarse-grained density variation $\delta n(x,t)$ translates directly into a coarse-grained variations $\delta y(x,t) \propto \delta n(x,t)$. The proportionality coefficient is purely geometrical: adding one atom to the edge results in its displacement by a lattice period in the climb direction $\delta y(x,t) =a'$ and also in a density change $\delta n(x,t)= 1/a$, where $a$ is the length of the unit cell along the core. Thus,
%\begin{eqnarray}
$\delta n(x,t)=\xi \delta y(x,t)$ with $\xi \equiv 1/aa'$.
%\label{n_y}
%\end{eqnarray}
This relation implies that for a superfluid dislocation the core displacement $\delta y$ is the conjugate variable to the superfluid phase $\varphi$. %, $[  \varphi(x), \xi \delta y(x')] = i \delta(x-x')$. 
The combined coarse-grained, low-energy effective action for superfluid and displacement degrees of freedom in the imaginary time description reads ($\hbar=1$)
\begin{eqnarray}
S= \int_0^{\beta}  d\tau\,  \int dx \left[ -i \xi y \dot{\varphi} + (\rho_s/2) (\partial_x\varphi)^2
- \mu \xi y\right] +S_d ,
\label{S}
\end{eqnarray}
where the purely dislocation part of the action,  $S_d$, is taken in the form of the Granato-L\"ucke string subject to Peierls potential \cite{Granato,Kosevich}:
\begin{eqnarray}
S_d =\int_0^{\beta} \! d\tau\, \int dx \left[ \frac{n_1v^2_d}{2}(\partial_x y) ^2 - u\cos\left(\frac{ 2\pi y}{a'}\right) \right],~~
\label{S_SG}
\end{eqnarray}
with $n_1$ being the linear mass density of the core, $v_d$ standing for speed of sound along the string determined by shear modulus $G$: $v^2_d \approx G/n_1$, and $u$  denoting the strength of Peierls potential. In Eq.(\ref{S_SG}), the kinetic energy $\propto \dot{y}^2$ is neglected in the low energy limit under the consideration. Full quantum mechanical description of the system based on calculating the partition function $\int \, Dy \, D\varphi \exp(-S)$ will be presented elsewhere.

Apart from the Peierls term $\propto u$ (not to be confused with the sine-Gordon term where the argument would be $\propto \int^x  y(x') dx'$),  the quantized action (\ref{S})-(\ref{S_SG}) is a standard harmonic $(1+1)$-dimensional action. A renormalization-group analysis, similar to the one given in Ref.~\cite{no_rough},  shows that, at $T=0$, the Peierls term has scaling dimension ${\rm dim}[u]=2$  regardless of the parameters of the system, even if the long-range deformation potential forces are ignored.  This means that the Peierls barrier
is relevant at $T=0$ and always leads to a finite gap $\Delta$ for the climb motion, i.e. the dislocation in its groundstate is smooth.
In such a state, the  cosine term can be expanded in powers of $ y$ around some equilibrium position $y_m= ma',\,\, m=0,\pm 1, \pm 2, ...$.
Accordingly, in the low-energy limit---when the gradient in the action (\ref{S_SG}) can be ignored---the action (\ref{S}) reduces to the standard 1D superfluid action \cite{Haldane}
\begin{eqnarray}
S_1= \int_0^{\beta}  d\tau\,  \int dx \left[ -i  \xi y \dot{\varphi} + {\rho_s\over 2} (\partial_x\varphi)^2 - \mu \xi y + {g\over 2} y^2 \right] ,~~
\label{S_1}
\end{eqnarray}
with $g=u (2\pi)^2/a'^2$. This action describes superfluidity with speed of sound $v_1=\sqrt{\rho_s g}/ \xi \propto \sqrt{\rho_s u}$ and also a finite climb in response to $\delta \mu$: $ \delta y = \xi \delta \mu  /g$.

With increasing $T$, thermally excited jogs and kinks render Peierls potential less and less relevant, so that  eventually it can be ignored.
In this limit, the dislocation becomes rough, that is, similar to a free string \cite{Granato}, and the spatial gradient in Eq.~(\ref{S_SG}) should be taken into account. The effective action (\ref{S}) then becomes
\begin{eqnarray}
S_2= \int_0^{\beta} d\tau\,  \int dx  \left[ -i  \xi y \dot{\varphi} + \frac{\rho_s}{2} (\partial_x\varphi)^2  + \frac{n_1v^2_d}{2}(\partial_x y) ^2 - \mu \xi y\right]. ~~
\label{S_2}
\end{eqnarray}
 Eq.(\ref{S_2}) predicts  an  {\it extremely strong} quasi-static climb response: $ \partial_x^2 \delta y \propto - \delta \mu$ determined by the length of a free dislocation segment $L$ (the cross-linking distance in the network), so that a typical displacement $\delta y \propto L^2 \delta \mu$. This implies that the resulting specific compressibility is {\it independent} of the dislocation density $\approx 1/L^2$, provided the network is uniform over the whole sample. Indeed, the added amount of atoms per each "elementary" cube of the side $L$ is $\sim a L\delta y \propto L^3 \delta \mu$. Thus, the added fraction of atoms per unit volume is independent of $L$.

 The superfluid component also demonstrates an unusual behavior. The equation of motion for small oscillations reads
\begin{eqnarray}
\ddot{\varphi} - \eta\,  \partial_x^4 \varphi =0, ~~~~~~ \eta \equiv \frac{\rho_s n_1 v_d^2}{\xi^2} ,
\label{quart}
\end{eqnarray}
meaning that the spectrum of superfluid excitations is not sound-like anymore. It is described by a {\it quadratic dispersion} $\omega = \sqrt{\eta} q^2$, where $q$ is the momentum along the dislocation line. Full quantum mechanical description of the crossover from the regime (\ref{S_1}) to (\ref{S_2}), (\ref{quart}) in line with the approach of Ref.~\cite{no_rough} will be presented elsewhere.
Here we point out two qualitative predictions of the model (\ref{S})-(\ref{S_SG}):  (i) suppression of the climb at $T< \Delta$, and (ii) dramatic softening of superfluid phonons at $T>\Delta$.

\begin{figure}
\centerline{\includegraphics[angle =0,width=0.9\columnwidth]{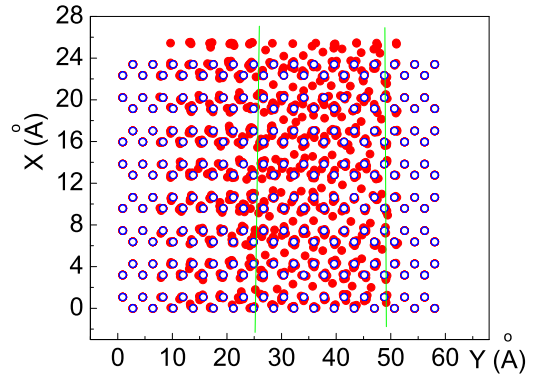}}
\vspace{-0.5cm}
\caption{(Color online)
Columnar view of a typical MC configuration along the C-axis: filled red dots show atomic positions; open blue circles indicate an ideal lattice; vertical solid green lines mark positions of the partial cores. The {\it fcc} fault is between these two lines. The superfluidity occurs along the green lines.
}\label{XY}
\end{figure}

\begin{figure}
\centerline{\includegraphics[angle =0,width=0.9\columnwidth]{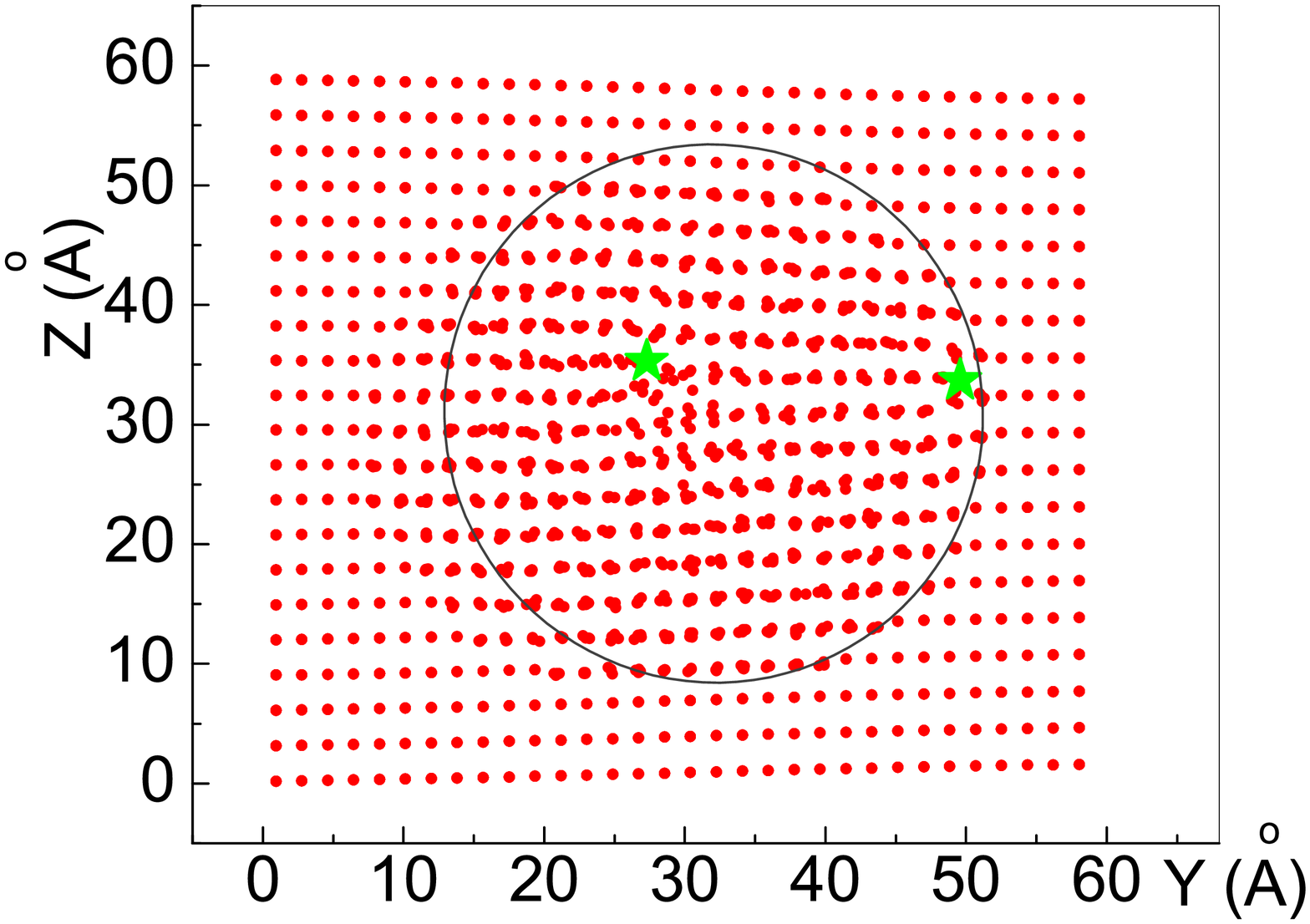}}
\vspace{-0.5cm}
\caption{(Color online)
Columnar view of the same MC configuration (as in Fig.~\ref{XY}) along the cylinder axis. The circle marks the simulation cell where particle positions have been updated.
The superclimb occurs in the``horizontal" plane.
Green stars mark positions of the two partials.
}\label{ZX}
\end{figure}

{\it Numerical results}. Our {\it ab initio} Monte Carlo (MC) simulations were based on the worm algorithm \cite{worm}.  The most important numerical finding of the present study is that edge dislocations with Burgers along the {\it hcp} axis 
have superfluid cores in solid {\he4}. Our example is based on the dislocation with the core
along the X-axis (and Burgers vector along the Z-axis).
Since the {\it hcp} structure has two atoms in the unit cell, two extra half-planes are involved.
Figures~\ref{XY}-\ref{ZX} show snapshots of atomic positions in a typical MC configuration, along C-axis and along the core. Particles outside the circle, Fig.~\ref{ZX}, were pinned to their classical lattice positions
and provided boundary conditions for the simulation cell.
The studied dislocation splits into two partials with the {\it fcc} fault forming in between \cite{Hull}.
The splitting is so large that it does not fit the simulation cell since one of the partials has moved
all the way to the cell boundary. A direct simulation of the {\it fcc} fault yielded an unmeasurably small
(within our accuracy) fault energy $<$ 0.1K/atom, meaning that the splitting (proportional to the
inverse of the fault energy) is indeed expected to be as
large as $\geq 150-300$\AA ~. Correspondingly, physical properties of the both partials are essentially
independent from each other.

Under these circumstances we performed extensive simulations of a single partial attached to the fault.
The rectangular simulation cell contained from 600 to 3400 particles with periodic boundary
conditions along the core. In perpendicular directions a boundary of pinned \he4
atoms surrounding a cylinder of radius $R$ provided the necessary boundary conditions for the
simulated sample of solid \he4 containing the partial dislocation at the center and the fault
extending in the positive Y-direction. Depending on $R$, the number of actually simulated particles
varied from 270 to 1700. Superfluid properties were detected by observing winding exchange cycles
along the cylinder axis (X-axis). The core response to changes in $\mu$  has been studied
by tracing the position of the maximum $Y(\mu)$ of the columnar superfluid density map in the
(Y,Z) plane \cite{methods}.

\begin{figure}
\centerline{\includegraphics[angle = 0,width=0.9\columnwidth]{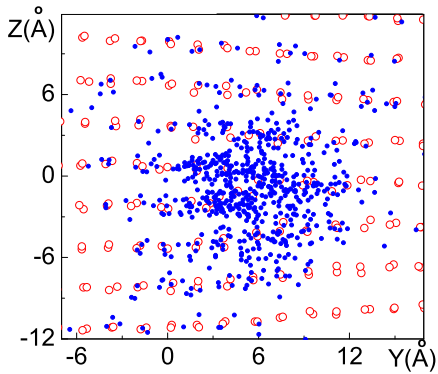}} %{fig3.eps}}
\caption{(Color online)
A columnar snapshot of atomic positions (open red dots) in the vicinity of the partial core (at the center)  superimposed
with the map (solid blue dots) of winding exchange cycles responsible for superfluid properties along the core, $X$-axis.
Note that (i) the map extends over several unit cells ($\approx 3.67$\AA~),
and (ii) it has no visible structure
in the Y-direction, implying negligibly small Peierls potential for climb  
at the simulated temperature.
}\label{deriv}
\end{figure}

The core position exhibited strong continuous response to variations of $\mu$.
The slope $dY/d\mu $ was larger in bigger cells indicating that at the simulated temperature $T=0.5~K$
it is controlled by the image forces provided by the boundary conditions.
At fixed $\mu$, the configuration-to-configuration fluctuations of the core position
were as large as several unit cells. Remarkably, the exchange-cycle map does not show any visible
modulation with the lattice period in the Y-direction (while the structure in the Z-direction is clearly seen), see Fig.~\ref{deriv}, meaning that the core
is loosing its crystalline structure locally and the Peierls potential in the climb direction 
is negligible under the simulated conditions. A systematic numeric study of the Peierls gap effects emerging
at much lower temperatures and in larger system sizes remains a major computational challenge.

%%%%%%%%%%%%%%%%%%%%%%%%%%%%%%%%%%%%%%%%%%%%%%%%%%%%%%%%%%%%%%%%%%%%%%%%%%%%%%%%%%%%%%%%%%%%%%%%%%
%%%%%%%%%%%%%%%%%%%%%%%%%%%%%%%%%%%%%%%%%%%%%%%%%%%%%%%%%%%%%%%%%%%%%%%%%%%%%%%%%%%%%%%%%%%%%%%%%%
%%%%%%%%%%%%%%%%%%%%%%%%%%%%%%%%%%%%%

%%%%%%%%%%%%%%%%%%%%%%%%%%%%%%%%%%%%%
%%%%%%%%%%%%%%%%%%%%%%%%%%%%%%%%%%%%%%%%%%%%%%%%%%%%%%%%%%%%%%%%%%%%%%%%%%%%%%%%%%%%%%%%%%%%%%%%%%

Crucial data can be obtained experimentally with the ``UMass sandwich"  setup, that potentially allows one to work at $T$ of few tens of $mK$ \cite{Hallock_private}. Since the quantity of interest is the isochoric compressibility (as a function of $T$), one can use the {\it superfluid syringe} experimental protocol, when both Vycor electrodes are being operated at one and the same chemical potential and are used exclusively to inject atoms into the solid, rather than to induce a DC flow. Such measurements near $400mK$ have been already done \cite{Ray3}.

Summarizing, we present strong {\it ab initio} evidence and a coarse-grained analytic description of the climbing of an edge dislocation in solid \he4, assisted by superfluidity of its core. This phenomenon yields a natural microscopic interpretation for the effect of anomalous isochoric compressibility accompanying superflow in the experiment by Ray and Hallock.
Theoretically, we argued that at low $T$, the superclimb, and, correspondingly,  the effect of anomalous isochoric compressibility, must be suppressed due to a crossover to a smooth dislocation. Experimental observation of the suppression, feasible within the ``UMass sandwich" setup, might yield strong support for the proposed scenario bridging ``quantum metallurgy" and supersolidity. The superclimb effect can also lead to high mobility of small dislocation loops (with Burgers vectors along C-axis) made of one partial surrounding an {\it fcc} fault. Such loops could be plenty in real samples (cf. \cite{Hull}), and implications of their presence are yet to be investigated.

%%%%%%%%%%%%%%%%%%%%%%%%%%%%%%%%%%%%%%%%%%%%%%%%

The authors are grateful to R. Hallock, D. Schmeltzer, and M. Troyer for stimulating discussions. We also thank P. Corboz for initial assistance.
This work was supported by the National Science Foundation under Grants Nos. PHY-0653183 and PHY-0653135,CUNY grants and the Swiss National Science Foundation.
Simulations were performed on Brutus (ETH Zurich), Typhon and Athena (CSI), and Masha (UMass) Beowulf clusters.


\begin{thebibliography}{99}

\bibitem{Andreev69} A. F. Andreev and I. M. Lifshitz, Sov. Phys. JETP {\bf 29}, 1107 (1969);
%\bibitem{Andreev69} Andreev, A. F. \& Lifshitz, I. M. Quantum theory of defects in crystals. {\it Sov. Phys. JETP} {\bf 29}, 1107 (1969).
%\bibitem{Thouless}
D. J. Thouless %,The flow of a dense superfluid, Ann. Phys. {\bf 52}, 403-427(1969)
Ann. Phys. {\bf 52}, 403(1969);
%\bibitem{Chester70}
G. V. Chester, Phys. Rev. A {\bf 2}, 256 (1970).
%\bibitem{Chester70} Chester, G. V. Speculations on Bose-Einstein condensation and quantum crystals. {\it Phys. Rev. A} {\bf 2}, 256 (1970).

\bibitem{KC} E. Kim and M.H.W. Chan, Nature {\bf 427}, 225
(2004); Science {\bf 305}, 1941 (2004).
%\bibitem{KCNature} E. Kim and M. H. W. Chan, Nature {\bf 427}, 225 (2004).
%\bibitem{KCNature} Kim, E. \& Chan, M. H. W. Probable observation of a supersolid helium phase. {\it Nature} {\bf 427}, 225-227 (2004).
%\bibitem{Rittner06}
A.S.C. Rittner and J.D. Reppy, Phys. Rev. Lett. {\bf 97}, 165301 (2006); {\it ibid} {\bf 98}, 175302 (2007);
%\bibitem{Rittner06} Rittner, AnnSophie C.  \& Reppy, J. D.  Observation of classical rotational inertia and nonclassical supersolid signals in solid ${}^4$He below 250 mK. {\it Phys. Rev. Lett. } {\bf 97}, 165301 (2006).
%\bibitem{Aoki}
Y. Aoki, {\it et al.}, %J. C. Graves, and H. Kojima,
Phys. Rev. Lett. {\bf 99}, 015301 (2007);
%\bibitem{Aoki} Aoki, Y. ,  Graves, J.C., \& Kojima, H. Oscillation frequency dependence of nonclassical rotation of solid ${}^4$He. {\it Phys. Rev. Lett. } {\bf 99}, 015301 (2007).
%\bibitem{Shirahama}
M. Kondo, {\it et al.}, %S. Takada, Y. Shibayama, and K. Shirahama,
J. Low temp. Phys. {\bf 148}, 695 (2007);
%\bibitem{Shirahama} Kondo, M. , Takada, S.  Shibayama, Y. \&  Shirahama, K. Observation of nonclassical rotational inertia in bulk solid ${}^4$He. {\it J. Low temp. Phys.} {\bf 148}, 695-699 (2007).
%\bibitem{Kubota}
A. Penzev, {\it et al.}, %Y. Yasuta, and M. Kubota,
J. Low temp. Phys. {\bf 148}, 677 (2007).
%\bibitem{Kubota} Penzev, A.,  Yasuta, Y. \& Kubota, M., Annealing effect for supersolid fraction in ${}^4$He. {\it J. Low temp. Phys.} {\bf 148}, 677 (2007).

\bibitem{meisel} M.W. Meisel, Physica B {\bf 178}, 121 (1992);
%\bibitem{meisel} Meisel, M. W. Supersolid ${}^4$He: an overview of past searches and future possibilities. {\it Physica B} {\bf 178}, 121-128 (1992).
%\bibitem{Greywall77a}
D.S. Greywall, Phys. Rev. B {\bf 16}, 1291 (1977);
%\bibitem{Greywall77a} Greywall, D. S. Search for superfluidity in solid ${}^4$He. {\it Phys. Rev. B} {\bf 16}, 1291 (1977).
%\bibitem{Day}
J. Day, and J. Beamish,  Phys. Rev. Lett. {\bf 96}, 105304 (2006);
%\bibitem{Day} Day, J. \& Beamish, J. Pressure-driven flow of solid Helium. {\it Phys. Rev. Lett. } {\bf 96}, 105304 (2006).
%\bibitem{Reppy}
A.S.C. Rittner, {\it et al.}, %W. Choi, E.J. Mueller, and J.D. Reppy,
%Absence of Pressure-Driven Supersolid Flow at Low Frequency
arXiv:0904.2640.

\bibitem{Ray} M.W. Ray and R.B. Hallock, Phys. Rev. Lett. {\bf 100}, 235301
(2008).

\bibitem{Ray2} M.W. Ray and R.B. Hallock, Phys. Rev. B {\bf 79}, 224302
(2009).


\bibitem{Balibar} S. Sasaki, {\it et al.}, % R. Ishiguro, F. Caupin, H.J. Maris, and S. Balibar,
Science {\bf 313}, 1098 (2006).
%\bibitem{Balibar} Sasaki, S., Ishiguro, R.,  Caupin, F., Maris, H. J.  \& Balibar, S. Superlfuidity of grain boundaries and supersolidity. {\it  Science} {\bf 313}, 1098 (2006).

\bibitem{Balibar_review} S. Balibar and F. Caupin, J. Phys.: Cond. Matter {\bf 20}, 173201 (2008).
%\bibitem{Balibar_review} Balibar, S. \& Caupin, F. Supersolidity and disorder. {\it J. Phys. Cond. Mat. } {\bf 20}, 173201 (2008).

\bibitem{Shevchenko}
S.I. Shevchenko, Sov. J. Low Temp. Phys. {\bf 13}, 61 (1987).

\bibitem{Mott} Note an analogy with incompressibility of the phase of a Mott insulator in externally imposed lattices.

\bibitem{PS} N. Prokof'ev and B. Svistunov, Phys. Rev. Lett. {\bf 94}, 155302 (2005).
%\bibitem{PS} Prokof'ev, N. \& Svistunov, B. Supersolid state of matter. {\it Phys. Rev. Lett.} {\bf 94}, 155302 (2005).

\bibitem{Boninsegni07} M. Boninsegni, {\it et al.}, %A.B. Kuklov, L. Pollet, N.V. Prokof'ev, B.V. Svistunov, and M. Troyer,
Phys. Rev. Lett. {\bf 99}, 035301 (2007).
%\bibitem{Boninsegni07} Boninsegni, M.,  Kuklov, A. B., Pollet, L., Prokof'ev, N. V., Svistunov, B. V. \& Troyer, M. Luttinger liquid in the core of a screw dislocation in Helium-4. {\it Phys. Rev. Lett.} {\bf 99}, 035301 (2007).

\bibitem{no_rough}
D. Aleinikava, {\it et al.}, %E. Dedits, A.B. Kuklov, and D. Schmeltzer,
arXiv:0812.0983.

\bibitem{Dorsey} A. Dorsey, talk at Workshop ``Supersolid 2008", Trieste, ITCP, August 18-22, 2008.

\bibitem{Meyerovich} A. Meyerovich, private communication.

\bibitem{deGennes}
P.G. de Gennes, C.R. Physique {\bf 7}, 561(2006).

\bibitem{Biroli}
J.-P. Bouchaud and G. Biroli, C.R. Physique {\bf 9},1067 (2008).
%Quantum plasticity and dislocation-induced supersolidity



\bibitem{Hull} D. Hull and D.J. Bacon, {\it Introduction to Dislocations}, Butterworth-Heinemann,  4th Edition, 2007.


\bibitem{Hirth} J.P. Hirth and  J. Lothe, {\it Theory of Dislocations}, McGraw-Hill, 1968.

\bibitem{Kosevich}
A.M. Kosevich, {\it The Crystal Lattice: Phonons, Solitons, Dislocations, Superlattices}, Wiley, 2005.
%\bibitem{dislo} SG model for dislocation???

\bibitem{Beamish}
J. Day and J. Beamish, Nature {\bf 450}, 853 (2007).

\bibitem{Granato} A. Granato and K. L\"ucke, J. Appl. Phys. {\bf 27}, 583 (1956); {\it ibid}. 789 (1956).

%\bibitem{comment_sine_Gordon} Within our representation, the cosine argument in the  sine-Gordon term would be proportional to $\int_0^x y(x') dx'$,
%as opposed to $y(x)$ of the Peierls term.

\bibitem{Haldane}
F. D. Haldane, Phys. Rev. Lett. {\bf 47}, 1840 (1981).
%Effective Harmonic-Fluid Approach to Low-Energy Properties of One-Dimensional Quantum Fluids


\bibitem{worm} M. Boninsegni, {\it et al.}, %N. Prokof'ev, and B. Svistunov,
Phys. Rev. Lett. {\bf 96}, 070601 (2006); Phys. Rev. E {\bf 74}, 036701 (2006).
%\bibitem{worm} Boninsegni, M., Prokof'ev, N.,  \& Svistunov, B., Worm algorithm for continuous-space path integral Monte Carlo simulations. {\it  Phys. Rev. Lett. } {\bf 96}, 070601 (2006).

\bibitem{methods} The methods of addressing superfluid properties of dislocation core are essentially the same that were used previously for revealing the superfluidity of the screw dislocation along the C-axis in Ref.~\cite{Boninsegni07}.

\bibitem{Hallock_private} R.B. Hallock, private communication.

\bibitem{Ray3} M.W. Ray and R.B. Hallock, arXiv:0908.2591.

\end{thebibliography}
\end{document}